# Toward AI-Ready Medical Imaging Data

Milen Nikolov[1,*], Edilberto Amorim[2], J Harry Caufield[3], Nayoon Gim[4], Nomi L Harris[3], Jared Houghtaling[5], Xiang Li[6], Danielle Morrison[7], Anaïs Rameau[7], Jamie Shaffer[4], Hari Trivedi[8], Monica C Munoz-Torres[9,*]

[1]Sage Bionetworks, 2901 Third Ave. Suite 330, Seattle WA 98121, USA
[2]UCSF Weill Institute for Neurosciences, 2540 23rd St., #5217, San Francisco, CA 94110, USA
[3]Lawrence Berkeley National Laboratory, 1 Cyclotron Rd., Berkeley CA 94720, USA
[4]University of Washington School of Medicine, 1959 NE Pacific St., Seattle, WA 98195, USA
[5]Tufts University School of Medicine, 136 Harrison Ave., Boston, MA 02111, USA
[6]Massachusetts General Hospital, 55 Fruit St., Boston, MA 02114, USA
[7]Sean Parker Institute for the Voice, Weill Cornell Medicine, 240E 59th St, New York, NY 10022, USA
[8]Emory University School of Medicine, 100 Woodruff Circle, Atlanta, GA 30322, USA
[9]Department of Biomedical Informatics, University of Colorado Anschutz, 13001 E 17th Pl., Aurora, CO 80045, USA

\* **Co-corresponding Authors**: Milen Nikolov (milen.nikolov@sagebase.org) *and* Monica C. Munoz-Torres (monica.munoztorres@cuanschutz.edu)

# Abstract

Medical imaging data plays a vital role in disease diagnosis, monitoring, and clinical research discovery. However, the complexity of medical imaging data, along with site-specific variations in data acquisition and metadata handling, presents challenges for multi-institutional collaboration and downstream analysis. Biomedical data managers and clinical researchers must navigate a complex landscape of medical imaging infrastructure, input/output tools and data reliability workflow configurations, taking months to operationalize.

While standard formats exist for medical imaging data, standard operating procedures (SOPs) for data management - including end-to-end guidelines for organizing, verifying, anonymizing, and preparing data for downstream integration and reuse, with a view towards AI/ML analysis - are lacking. These data management SOPs are key for developing Findable, Accessible, Interoperable, and Reusable (FAIR) data, a prerequisite for AI-ready datasets.

The National Institutes of Health (NIH) Bridge to Artificial Intelligence (Bridge2AI) Standards Working Group members and domain-expert stakeholders from the Bridge2AI Grand Challenges teams developed data management SOPs for the Digital Imaging and Communications in Medicine (DICOM) format. This article focuses on DICOM since 1) all Bridge2AI medical imaging data is standardized and produced as DICOM files; 2) DICOM is the internationally recognized standard for medical imaging (ISO 12052), implemented across a number of modalities: X-ray, CT, MRI, ultrasound, ophthalmology; 3) DICOM is the most commonly-used medical imaging format, and is supported by a rich ecosystem of visualization and analysis tools.

We describe best practices for DICOM data and metadata reliability, applying to both static and video imaging modalities. We emphasize steps required for centralized data aggregation, validation, and de-identification, including a review of defacing methods for facial DICOM scans, anticipating adversarial AI/ML data re-identification methods. Data management vignettes based on Bridge2AI datasets include example parameters for efficient capture of a wide modality spectrum, from ultrasound and endoscopic videos through ophthalmology retinal scans.

# Introduction

The Bridge2AI program, funded by the National Institutes of Health (NIH), aims to accelerate the integration of artificial intelligence (AI) approaches into biomedical research by creating standardized, annotated, ethically sourced, and diverse AI-ready datasets. Its mission is to establish best practices for generating data that is not only technically robust but also equitable and transparent. As part of this initiative, four grand challenge projects – Precision Public Health, Salutogenesis, Clinical Care, and Functional Genomics – produce high-quality, large-scale datasets across a diverse set of modalities. These projects collectively seek to set new standards for how biomedical data is generated, shared, and leveraged to advance trustworthy AI solutions in health [1].

Central to many of these efforts is the use of Digital Imaging and Communications in Medicine (DICOM), the global standard for storing, transmitting, and managing medical imaging data [2]. All Bridge2AI grand challenges generating medical imaging data have converged on DICOM as a foundation for imaging data acquisition and processing. The Clinical Care grand challenge [3] leverages DICOM standards to capture and harmonize all imaging data acquired in an ICU setting (e.g. MRI, CT scans, ultrasound, X-Ray), linking the imaging data with physiologic and clinical data. Harmonized and linked data enables AI models to predict patient trajectories in critical care. To elucidate mechanisms of chronic diseases like diabetes and its complications, the Salutogenesis [4] challenge applies DICOM's ophthalmology specifications to harmonize retinal scans, resulting in the groundbreaking Artificial Intelligence Ready and Equitable Atlas for Diabetes Insights (AI-READI) dataset, the first publicly available standards-compliant DICOM retinal imaging dataset [5]. The Precision Public Health [6] challenge uses DICOM for brain MRIs and CT scans as well as laryngoscopy videos, connected with voice and omics data, resulting in a multi-modal AI training dataset.

DICOM data management best practices are essential to Bridge2AI. Ensuring accurate metadata enables medical imaging data interoperability and downstream data integration across diverse healthcare systems. DICOM provides the necessary infrastructure for linking imaging data to metadata, clinical context, and outcomes – a critical component in making data Findable, Accessible, Interoperable and Reusable (FAIR) and advancing multimodal biomedical AI. The Bridge2AI Standards, Practices, and Quality Assessment Working Group developed this article to guide the implementation of DICOM best practices across Bridge2AI and beyond, ensuring that the resulting end-to-end workflows and datasets 1) enable future biomedical AI research; and 2) provide utility to biomedical data managers, bioinformaticians and clinical researchers who currently navigate a complex ecosystem of medical imaging data infrastructure without a concise, user-friendly guide.

Effective data lifecycle pipelines require an integrated multi-stage workflow that spans tasks from initial data capture and characterization through downstream cloud storage and integration with health records. To support consistent, reproducible, and scalable use of medical imaging data in research, this document presents a detailed *operational* framework that covers a full imaging data lifecycle workflow, along with tactical resources and examples for each of the workflow stages. Each stage is described with definitions, actionable recommendations and, where applicable, examples from real-world datasets, tooling options, and references to published literature or standards.

We organize the imaging data lifecycle strategy in seven stages.

**Stage 1. Data extraction and metadata characterization:** imaging data must be extracted from clinical repositories such as Picture Archiving and Communication Systems (PACS) [7] that are used for storing and accessing imaging data. Often guided by tools like data landscape assessment forms (see Appendix I), this data management stage lays the foundation for subsequent verification and transformation stages.

Once extracted, *reliable imaging data must be complete, standardized, and biologically realistic*. Data reliability workflows include practical techniques for spotting outliers, verifying imaging protocols, and identifying structurally invalid DICOMs. These workflows comprise stages 2 - 7.

**Stage 2. File integrity verification:** file corruption due to network errors, media decay, or system failure can degrade research datasets. This stage includes checksum-based integrity monitoring using SHA-256 or MD5, in line with NIST best practices and the RSNA Image Sharing initiative [8]. Automated checksum workflows are increasingly common in imaging repositories (e.g., the Extensible Neuroimaging Archive Toolkit (XNAT) [9], OpenNeuro [10]) and are now considered a best practice in bioinformatics pipelines. These guidelines include simple Python example snippets for using checksums or hashes (e.g., SHA-256 [11]) to ensure data has not been altered during storage or transmission.

**Stage 3. Data completeness and conformance checks:** it is critical that mandatory DICOM fields (e.g., PatientID, Modality, StudyInstanceUID) are populated and consistent within and across series, as inconsistencies can corrupt longitudinal analyses and patient-level data merging. Methods such as header consistency audits and intra-study UID checks are used in this stage, building on validation protocols used in the Clinical Care and Salutogenesis Grand Challenges. These approaches align with FDA guidance [12] on imaging data in regulatory submissions, among other standards.

**Stage 4. Metadata tag validation** against published DICOM standards (e.g., PS3.3 2024 edition by NEMA [13]) is critical to ensure semantic interoperability and prevent downstream software errors. Errors in DICOM tag structure or coding syntax often result in failed reads or loss of semantic fidelity [14]. We outline how to catch and correct such issues early and provide example code snippets, using tools such as `pydicom` [15], DVTk [16], and `dicom3tools` [17] to enable automated conformance checking.

**Stage 5. Image quality and pixel data validation** includes assessing the integrity of pixel arrays, compression methods, and spatial attributes like pixel spacing and slice thickness. Image fidelity is critical for downstream AI applications and quantitative analysis. We provide example code snippets for testing pixel readability (e.g., via `pixel_array` in `pydicom`), spatial resolution, and compression formats.

**Stage 6. Metadata consistency and correctness** complements pixel validation by ensuring that unique identifiers (UIDs) are properly formed, demographic metadata is logically coherent, and timestamps are valid across study sessions. Metadata inconsistencies – including mismatched dates, repeated UIDs, or biologically implausible entries – undermine dataset integrity. We summarize best practices from Integrating the Healthcare Enterprise (IHE) [18] and OHDSI [19] initiatives, providing example checks for identifier uniqueness and demographic coherence. These checks are essential for avoiding duplicate records or incorrect linkage between patients and their imaging histories.

**Stage 7. Pixel-level and metadata de-identification**, such as defacing to obscure facial features in MRI scans, must be performed to protect patient privacy as datasets are prepared for broader use. In modalities like MRI and CT, facial structures may be reconstructed from pixel data, violating patient privacy. Tools like RSNA CTP [20], `mri_deface` and `pydeface` are often employed to meet privacy standards. These steps address ethical obligations and legal requirements under regulations such as HIPAA and GDPR. The methods described in this article are aligned with these regulations.

We cover all stages in this DICOM data lifecycle strategy in more detail below.

# Stage 1: Data Extraction and Metadata Characterization

Medical imaging research relies on accurate, standardized acquisition and archiving of images **from the point of capture onward**. Whether working with static images (e.g., CT, MRI, X-ray, ophthalmic photographs) or multi-frame video sequences (e.g., ultrasound cine loops, endoscopic video), the DICOM standard underpins data capture workflows. Reproducible data management practices – beginning at image acquisition – are essential to ensure that downstream analyses, AI/ML development, and multi-center studies can interoperate effectively. In practice, this means 1) identifying all relevant imaging sources and workflows (PACS, modality workstations, vendor systems, research archives); 2) extracting image pixel data and rich metadata in a modality-sensitive manner; and 3) characterizing and standardizing this data according to accepted protocols (e.g., DICOM, IHE, HL7/FHIR [21] to capture patient encounter information).

Early characterization and quality control of imaging metadata reduces errors and inconsistencies that can otherwise hinder interoperability and data reuse. Embedding metadata standards and provenance checking at acquisition dramatically lowers downstream curation costs and improves reproducibility in imaging research [14].

Compliance with interoperability standards (IHE, DICOM conformance) should guide data extraction. The following sections review data extraction operations and tools interoperable with these standard frameworks.

## A. Overview of Imaging Modalities, Data and Metadata Characteristics

Imaging data varies widely by modality. CT and MRI produce 3D volumetric series; X-ray yields 2D projection images; ophthalmic modalities - fundus photos, optical coherence tomography (OCT), OCT angiography (OCTA) - produce specialized scans of the retina [5]; and many modalities - e.g. ultrasound, video endoscopy - output multi-frame sequences or "cine" clips. Each modality has distinct DICOM Service Object Pair (SOP) Classes and key metadata elements that need to be captured. Figure 1 illustrates a typical mapping between DICOM SOPs and metadata.

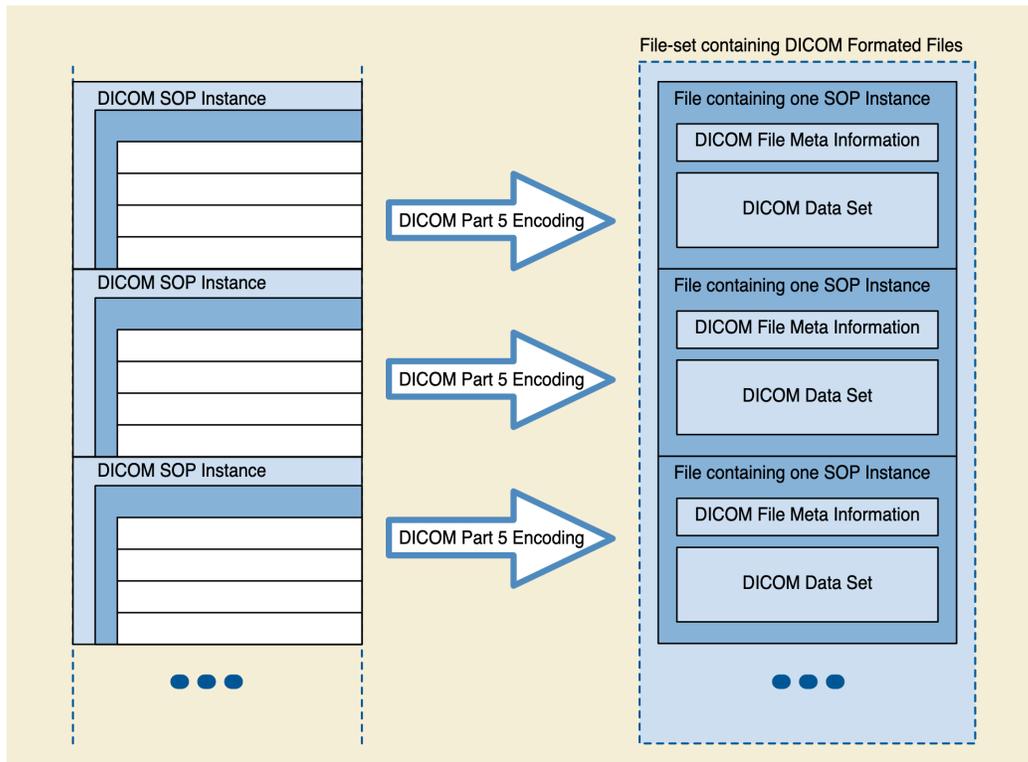

**Figure 1.** Typical map between DICOM Service Object Pair (SOP) instances and file data and metadata (image reproduced from [22]). In DICOM, a Service–Object Pair (SOP), consists of (a) an Information Object Definition (IOD) defining modality of data captured: a CT image, an ultrasound video, etc.; and (b) a DICOM Service defining what you can do with an object: store, query, retrieve it, etc.. The DICOM standard pre-defines a set of SOP classes: each class represents a specific combination of object type and service. Each SOP class has a unique ID (UID). Each DICOM file header contains instances (references) to the relevant SOP class (**left** in the figure), through the class UIDs, defining all standard operations that can be performed on the file.

The DICOM standard is organized into numbered parts. For example, Part 5, "Data Structures and Encoding (ISO 12052-5)", defines how DICOM data is physically represented in files for each SOP instance: how all attributes (tags) and pixel data are encoded in bytes in a standard way (**right** in the figure). This standardized encoding ensures interoperability across scanners, PACS and downstream processing tools, discussed in this article, e.g. `pydicom`, DCMTK, Orthanc.

A typical DICOM file-structure is flexible and can accommodate various modalities. Across all modalities, **metadata capture** is paramount. DICOM headers embed demographic and study information (Patient ID, Study/Series Instance UID, Acquisition Date/Time, Modality, etc.), and modality-specific acquisition parameters. Ensuring that common fields (e.g., patient name/ID, study description) are correct at acquisition helps downstream linking and anonymization, as discussed in sections below. For example, missing or inconsistent *Body Part Examined* or *Laterality* tags can complicate downstream queries; and mapping free-text tags to codes (e.g., mapping *Body Part Examined* to SNOMED) greatly improves FHIR interoperability.

Table 1 lists modality-specific transfer syntax, DICOM SOP Classes, and their mapping to DICOM file data and metadata. We assume the reader has familiarity with the basic structure and purpose of the general DICOM file format fields, as described in the DICOM standard [23].

**Table 1.** DICOM modalities and key metadata fields.

| Imaging Modality | Description | Key Metadata | Extraction Challenges | Special Considerations |
|---|---|---|---|---|
| CT and MRI | Multi-slice "stack" images, hundreds to thousands of slices per study | Image Position, Image Orientation, Pixel Spacing, Slice Thickness, Series Description, Series Instance UID, Instance Number | Collecting and ordering all instances for each series by Instance Number | Preserving special acquisitions (multi-echo, diffusion, spectroscopy) tags like Echo Time, Diffusion b-values, Sequence Name |
| X-ray (Radiography) | Single-image or few-image studies, 2D | View Position, Image Laterality, dose information, Series Description | Ensuring accurate laterality and projection labeling | View Position and Laterality can often be coded incorrectly, manual review may be necessary. |
| Ophthalmic Imaging (Photography, OCT, OCTA) | Single-image or Multi-frame scans of the retina | Laterality, Ophthalmic Image Resolution (Rows, Columns, Number of Slices), Camera/Device Serial Number, Scan Pattern | Handling large multi-frame DICOM files and associated nested DICOM tag information | Mapping codes to controlled vocabularies (e.g., SNOMED or RadLex), using standard storage classes (e.g., Ophthalmic Tomography 3D Retina Image Storage) |
| DICOM Video Sequences | Multi-frame video storage (e.g., ultrasound B-mode, color Doppler, endoscopic or | Frame Increment Pointers, Cine Rate, per-frame timestamps | Decoding video stream, re-packaging frames as needed | Preserving timing and image orientation tags, logging probe orientation or patient positioning, pixel de-identification is a |

| Imaging Modality | Description | Key Metadata | Extraction Challenges | Special Considerations |
|---|---|---|---|---|
| | laparoscopic videos) | | | challenge as anatomic information is often burned into the image |

**File organization** complements metadata fields. For example, DICOM files for radiology exams and other modalities are typically organized in a hierarchical folder structure (Figure 2), which can aid in identification of relevant images and checking for data-completeness, as follows: PatientID -> Accession Number (Exam ID) -> Series Instance UID -> DICOM files (named based on SOP Instance UID).

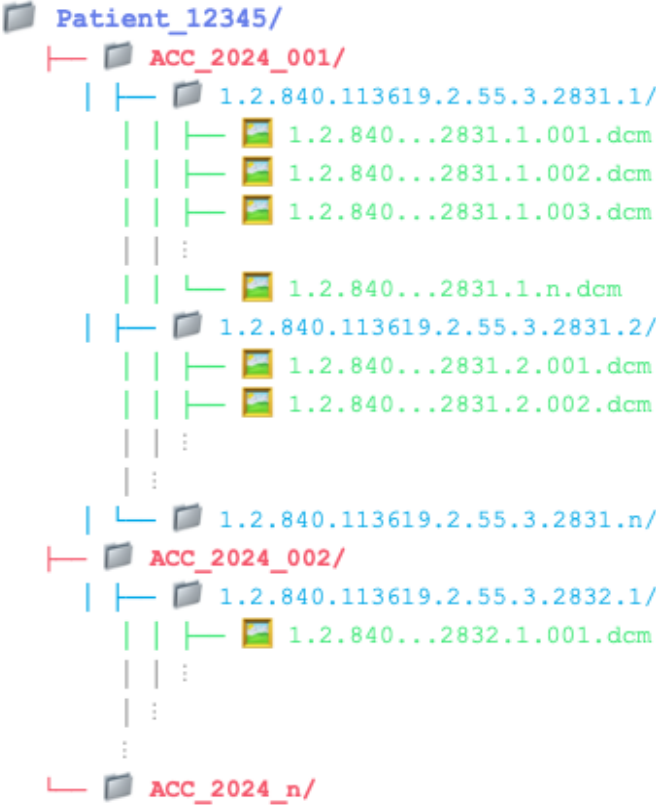

**Figure 2.** Example hierarchical folder structure for radiology DICOM image files. The patient (identified by their ID, Patient_12345) is at the top level, with a folder below that for each exam of that patient (ACC identifiers), etc.. Each exam folder contains corresponding DICOM files. A similar hierarchical structure is used in radiology and other modalities.

## B. Identifying and Documenting DICOM Imaging Sources

In a hospital or research network, imaging data sources can be characterized along four distinct dimensions.

**1. Physical and organizational source locations:** images are often distributed across clinical and departmental infrastructures. A research institution may maintain a central radiology PACS alongside local imaging archives; for example, a cardiology catheterization lab or an image store for an ophthalmology lab. Each department may enforce its own data governance rules, storage policies, and access controls, resulting in potential data silos.

**2. Database sources:** beyond departmental repositories, imaging data may be managed within institutional or research-specific databases and archives. These include frameworks like XNAT [9], an open-source imaging informatics platform widely used to manage, store, and share medical imaging data, particularly in neuroscience research. For each system, information important for downstream data management includes the vendor, software version, supported query mechanisms (e.g., DICOM C-FIND/C-MOVE [13], DICOMweb [24], or custom APIs); patient identifier handling; data retention; and anonymization capabilities.

**3. Data modalities sources and file formats:** different clinical domains (radiology, cardiology, ophthalmology, pathology, etc.) and respective imaging modalities (CT, MRI, ultrasound, endoscopy, OCT, etc.) often use distinct DICOM SOP Classes and data formats. Cataloging data modalities and file format sources is crucial to developing effective extraction and integration strategies downstream.

**4. Image acquisition systems:** imaging devices themselves, including CT scanners, MRI units, ultrasound machines, endoscope cameras, and OCT systems, produce data through vendor-specific software that may vary in DICOM compliance and export functionality. For example, some endoscopes may only export JPEG images rather than full DICOM objects.

A comprehensive **landscape assessment** should systematically capture information across these four dimensions. **An example landscape assessment capturing this information is provided in Appendix I.**

Collecting this landscape information early enables accurate planning and helps avoid unexpected challenges during data extraction and harmonization.

## C. DICOM Data Extraction

Once DICOM data sources have been catalogued, standardized data extraction protocols specific to the identified database source APIs, data modalities and imaging devices can be deployed. Popular protocols for extracting DICOM data include: 1) DICOM Query/Retrieve: using C-FIND and C-MOVE operations to bulk-transfer image series from PACS; 2) DICOMweb (WADO/QIDO): using RESTful WADO-RS or STOW-RS services to query and fetch images [13]; 3) Vendor APIs: some vendors provide database queries or regular dumps. The most common open source tools implementing standard DICOM data extraction protocols include:

- **DCMTK** [25] is a mature and efficient C/C++ toolkit providing command-line utilities and libraries for nearly all DICOM data capture standard services (C-STORE, C-MOVE, C-FIND, etc.). It is often installed on Linux/PACS servers. DCMTK can send/receive images and modify tags.

- **DICOMweb** (part of DICOM PS3.18) [13] defines RESTful services (QIDO-RS for query, WADO-RS for retrieve, STOW-RS for store) that allow web-based access to imaging data. Modern PACS and VNA systems often implement these services.

Once DICOM data is retrieved, image volume reconstruction often follows modality-specific workflows. Below we review key steps and configurations of image reconstruction in CT, MRI, X-rays, ophthalmology, video and multi-frame endoscopy modalities. We provide Bridge2AI grand challenge team vignettes including metadata configuration and workflow steps across these modalities.

For **CT and MRI data**, complete volume reconstruction requires assembling many slices within a series. Typically, DICOM extraction workflows group images by *Series Instance UID,* so that slices are not mixed across series. If a modality outputs multiple series (e.g., CT with and without contrast) or multi-phase scans, these typically appear as separate series. Tools like `dicomsort` or `pydicom` and downstream tools like the Insight Toolkit (ITK) [26] can organize and verify these series.

**Table 2**. Key operations in CT and MRI DICOM data reconstruction used by Bridge2AI teams.

| Operation | Workflow Tip |
| --- | --- |
| Frame Ordering | Use Instance Number or Image Position to sort slices correctly along the imaging plane. |
| Spacing and Orientation | Confirm that Pixel Spacing, Row/Column Orientation vectors, and Slice Thickness are present. If any vital geometry tags are missing, manual correction may be needed. |
| Multiframe Encapsulation | Some MRI scanners output a single multi-frame DICOM file encapsulating all slices. These files require unpacking. `pydicom` can read multi-frame objects by iterating over Pixel Data frames; DCMTK's `dcm2dcm` or ITK's GDCM also support multi-frame splitting. |
| MRI Sequence Details | Verify presence of protocol parameters from tags. These include TR, TE, Flip Angle, Echo Train Length, and others. |

**X-rays** yield one or few images per study. Depending on the vendor and institution, all X-ray images may appear within a single series, or the exam may contain multiple series with one image each. It is important to verify *Image Laterality* (Left/Right), *View Position, and Series Description* tags to ensure the correct images are being used for any desired workflow or training (e.g., craniocaudal vs. mediolateral projections).

Radiographic images may be stored anywhere from 8-bit to 16-bit images, although practically 12-bit is most common in modern systems. Some images are black-white inverted when stored; in these instances the *Photometric Interpretation* tag will indicate if the pixel intensities need to be inverted before conversion to another format such as PNG or JPG.

**Ophthalmic images** use DICOM Storage classes such as *Ophthalmic Photography 2D (e.g., fundus photo)*, *Ophthalmic Tomography (OCT, OCTA)*, and *Ophthalmic Axial Length*. These classes include unique tags (e.g., *Scene Illumination Color*, *Field of View Dimensions*, *Scanning Sequence*).

Often ophthalmic devices use custom DICOM tags (or private tags) for additional parameters. Extraction scripts should be robust to variation in these tag-sets. Capturing these data from the outset is important: including OCT/OCTA metadata in DICOM enables future AI analysis, but only if all critical fields - including proprietary ones are documented. It is good practice to log any non-standard tags encountered and consult device documentation. During extraction, one must use the correct DICOM SOP Class UID to retrieve data. For instance, OCTA is stored under *Ophthalmic Tomography 3D Angiography Image Storage* (SOP Class 1.2.840.10008.5.1.4.1.1.77.1.4). Tools like `pydicom` can be used to parse these tags. **Appendix II provides example vignettes** by the Bridge2AI's Salutogenesis grand challenge team, for configuring and working with these DICOM tags.

**DICOM video and multi-frame**, including cine loops, real-time sequences, and data extraction workflows must handle multi-frame transfer syntax. Picking the right video compression format and documenting respective parameters suitable for a given data modality and downstream application is important: some DICOM toolkits (e.g., `pydicom`) may require special handling to decompress MPEG or JPEG2000 frames. The DICOM attribute *Number of Frames* indicates how many images are packed. Each frame's position in time may be given by *Frame Increment Pointer* and related tags. When extracting data, a workflow is configured to save each frame as a separate image file (e.g. naming by frame index) or to keep the native multi-frame DICOM. Note that reading the latter needs to be supported by downstream tools. For example, tools like DCMTK's `storescu` (with `-multi-frame` option) or Orthanc [27] plugins can be scripted to retrieve all frames, and the GDCM [28] toolkit provides command-line utilities like `gdcmdump` to inspect these frames. If DICOM multi-frame data is converted to a video format (e.g. MP4), each DICOM frame is first decoded to a raw image, and then reassembled into an MP4 video. Frame timing tags enable temporal analysis downstream (e.g., in cardiac ultrasound studies); preserving these tags when converting to video formats is critical in these use cases. **Appendix III provides example vignettes by Bridge2AI's Precision Public Health grand challenge team**, including configuration options for capturing video DICOM for ultrasound, endoscopy and cine MRI.

Across all modalities, capturing provenance is a key part of DICOM data extraction. Provenance records which PACS or archive each image came from, the date of extraction, and the parameters used. Proper attribution of image source can be preserved in metadata or in a separate manifest, which aids auditing and reproducibility. There are various mechanisms to record provenance, or data lineage; however, this is outside the scope of these guidelines[1].

---

[1] The interested reader can find more about provenance standards starting with an open and popular standard implemented by various systems: W3Prov.

# Data Reliability Review

Once data is captured from imaging devices and source PACS, it is **critical** to rigorously review data reliability and ensure data quality. These activities take place in stages 2-6.

A robust data-quality framework for medical imaging comprises three pillars: **completeness, conformance, and biological plausibility**. These data reliability pillars are aligned with FAIR Data principles and are critical for AI-ready data [1].

Data **completeness** requires that all expected data files and metadata fields are present. For a given study, this includes every image, slice, and associated DICOM object.

Data **conformance** requires adherence to standard specifications. Each DICOM file must conform structurally to the DICOM standard (PS3.3–PS3.6) and follow the protocol workflow: expected sequences of images, tags, and values. Conformance checks catch malformed DICOMs or values not matching their defined Value Representation (VR).

Data **plausibility** requires that DICOM data and metadata content makes sense medically and technically. For example, patient age and sex match the study type; image geometry is within realistic bounds. Examples of implausible data might include a newborn with a mammogram, or a CT slice thickness of 0.

Often data reliability checks include outlier detection, protocol verification and structural validation. While some outlier detection can be done using DICOM metadata - for example, a *Series Description* that mentions 'Knee' in a chest X-ray exam is obviously an error - often these are only caught using manual or AI-based review of images. Protocol verification involves checking that DICOM metadata (e.g., *SequenceName*, *SliceThickness*, *RepetitionTime*, *PixelSpacing*) fall within expected ranges for the declared exam type. For example, an abdominal CT normally has a slice thickness of about 1.5 mm; a slice thickness of 1500 mm would be suspicious. Since outlier detection workflows and algorithms are highly-specific to study types, here we focus instead on typical file integrity verification, data completeness and consistency checks as part of structural data reliability requirements.

## *Stage 2: File Integrity Verification*

Medical images are often large and long-lived. Silent corruption (bit-rot, storage glitches) can render them unusable or misleading. Best practice for avoiding silent corruption requires computing and tracking cryptographic checksums during data capture and verifying these checksums regularly. NIST recommends using secure hashes like SHA-256 (FIPS 180-4) for integrity checks. For example, a repository might compute each DICOM file's SHA-256 at upload and store it in the database. Any later change in the file will change the hash, signaling corruption. MD5, though weaker for security, is still widely used for integrity checks because it's faster; many toolkits maintain both MD5 and SHA-256 checksums to balance speed and strength. The snippet below provides a basic checksum compute routine.

```python
""" Python code snippet for computing checksums of dicom images """
import hashlib

def checksum_sha256(file_path):
    """Compute SHA-256 checksum of a file."""
    h = hashlib.sha256()
    with open(''path_to_file', 'rb') as f:
        for block in iter(lambda: f.read(65536), b''):
            h.update(block)
    return h.hexdigest()

print(checksum_sha256('example_DICOM_image.dcm'))
```

In practice, most institutions and research networks rely on off-the-shelf file integrity solutions. The popular XNAT repository toolkit includes the MD5 comparator: XNAT's NRG Selenium pipelines provide a FileComparator - `MD5_Comparator` - that reads a file and compares its MD5 to an expected value. If users have a directory of DICOMs on disk and an accompanying manifest of SHA-256 checksums, they could write a script to recompute each file's hash and compare it with the original periodically. If any mismatch is found, the file has been corrupted.

OpenNeuro operationalizes similar checksum on upload: DICOM files are hashed as they enter the system and are checked periodically, automatically.

## Stage 3: Data Completeness and Conformance Checks

The checks in this stage verify that all expected data files and metadata fields are present and conform to standards. Structurally invalid DICOMs include incomplete files (e.g. truncated transfer syntax), wrong Value Representation (VR) encoding, or files lacking required tags. To detect structurally invalid DICOMs, tools like `pydicom` can be used to read a dataset. If mandatory tags (e.g., *StudyInstanceUID*, *PixelData*) are missing or corrupted, the read will fail or leave errors. Automated tools such as DCMTK's `dciodvfy` flag missing or illegal values; and DVTk can be used to validate object conformance. Code snippets showing examples for these validation tasks are provided in this and following sections.

Within and across studies, key DICOM identifiers and fields must be complete and consistent. Two common fields to audit are *PatientID* and *StudyInstanceUID*. Every DICOM in a study should carry the same *StudyInstanceUID* (to link series and images together) and the same *PatientID* (to group by subject). Missing or blank *PatientIDs* can break this linkage. For example, if an entire series of images accidentally omits *PatientID*, the images in the series may not be linked to relevant demographics data. It is thus prudent to scan each study folder and ensure all files share the same *StudyInstanceUID* and *PatientID*. The following example illustrates using `pydicom` to flag any study directory where more than one *StudyInstanceUID* is used. *PatientIDs* could be checked similarly.

```
# Pydicom codesnippet verifying all images within a study folder share a study ID
```

```python
import os, pydicom

study_dir = 'Example Study Name'
uids = set()
for fname in os.listdir(study_dir):
    ds = pydicom.dcmread(os.path.join(study_dir, fname), stop_before_pixels=True)
    uids.add(ds.StudyInstanceUID)
if len(uids) > 1:
    print("Inconsistent StudyInstanceUIDs found:", uids)
```

In Bridge2AI, the Clinical Care grand challenge team has implemented a DICOM to OMOP [29] [30] interface, as described in [31], to robustly link DICOM data and clinical encounters. The team maps patient header metadata - relevant to a local data collection site - to standard OMOP header metadata. This patient-linking guide identifies accession numbers for specific image tests, simplifying image test mapping to the primary encounter of interest for a multimodal data pull.

There are many other solutions for bridging to electronic health records, clinical data models, and DICOM images. For example, HL7 FHIR (R4 and newer) includes the *ImagingStudy* resource, which represents a DICOM study and its series/instances. Existing pipelines (e.g. [32]) can convert PACS DICOM metadata into FHIR *ImagingStudy* resources compliant with the HL7 core profile. This approach allows integration of imaging metadata into research data warehouses that already use FHIR (or other HL7-based) schemas.

Repeated UIDs pose another data management challenge: merging datasets may create *StudyInstanceUID* conflicts between distinct studies. One should ensure UIDs are unique. When combining data from multiple sources, it is possible for two different scans to, erroneously, have the same *SOPInstanceUID*. A simple check accumulates all *SOPInstanceUIDs* in a dataset: if any UID appears more than once in different files, that indicates a conflict, as illustrated in the code snippet below. If duplicates are found, the dataset manager must resolve which entry is correct (often by consulting acquisition logs) and assign a new UID to one copy.

```python
""" Pydicom code snippet verifying all scans have unique IDs across studies"""

uids = {}
for file in os.listdir('Example_AllStudies'):
    ds = pydicom.dcmread(os.path.join('Example_AllStudies', file), stop_before_pixels=True)
    uid = ds.SOPInstanceUID
    uids.setdefault(uid, []).append(file)
for uid, files in uids.items():
    if len(files) > 1:
        print(f"Duplicate UID {uid} in files: {files}")
```

Many archives generate new, globally unique UIDs, for example using MD5/SHA of the original UID, to avoid collisions.

These metadata consistency checks are usually complemented by auditing other fields: dates should follow logical order (e.g., *StudyDate* not before *PatientBirthDate*) and modalities should match planned protocols. Inconsistencies might reveal data entry errors or de-identification gaps. For instance, if two series in one study have different frame times or spatial dimensions when they were expected identical (e.g., two echoes of the same sequence), that could signal a corrupt or incomplete transfer. Applying conditional logic (e.g., "if *SpacingBetweenSlices* differs by more than 10% within a series, flag it") helps catch subtle anomalies. Tools like `pydicom` make it easy to script such checks, ensuring that each dataset is internally coherent before analysis or sharing. The Bridge2AI Clinical Care and Salutogenesis grand challenge teams, which aggregate multi-center imaging cohorts, emphasize internal coherence: all sites must use a unified identifier scheme and consistent demographic tagging so that merged data is queryable. At a minimum, data managers should verify that all scans have unique IDs. The `pydicom` library provides utilities for this, as illustrated above.

## Stage 4: Metadata Tag Validation

Tag validation is a critical step to catch malformed DICOM metadata. DICOM metadata tags must satisfy the standard's structural and syntactic rules. The DICOM standard defines a VR (Value Representation), VM (Value Multiplicity), and Type (1, 1C, 2, etc.), for each tag. For instance, the (0010,0020) *PatientID* configuration has VR=LO, VM=1, Type=2 and indicates that the field must be present, but can be empty. The *PixelSpacing* tag (0028,0030) has VR=DS, VM=2, indicating the field must contain two decimal strings for row and column spacing. A tag validation tool should check that data is conformant with these rules. Table 3 below provides a reference of critical DICOM tags, summarizing their standard definitions as of DICOM PS3.6.

**Table 3.** Critical tags in DICOM images that should be checked.

| Tag (Hex) | Name | VR | VM | Type | Description |
| --- | --- | --- | --- | --- | --- |
| (0010,0010) | PatientName | PN | 1 | 2 | Patient's full name (may be blank) |
| (0010,0020) | PatientID | LO | 1 | 2 | Patient identifier |
| (0020,000D) | StudyInstanceUID | UI | 1 | 1 | Unique ID for the study |
| (0020,000E) | SeriesInstanceUID | UI | 1 | 1C | Unique ID for the series |
| (0028,0010) | Rows | US | 1 | 1 | Number of rows (pixels) |
| (0028,0011) | Columns | US | 1 | 1 | Number of columns (pixels) |
| (0028,0030) | PixelSpacing | DS | 2 | 1 | Physical distance between pixels (mm) |

Libraries like `pydicom` allow custom checks. For example, the following snippet loads a DICOM file and verifies that each element's VR matches the dictionary:

```python
# Pydicom codesnippet checking the VR tag matches expected standard structure.
# Similar, logic could be used to check other tags

from pydicom import dcmread
from pydicom.datadict import dictionary_VR

ds = dcmread('Example_image.dcm')
for elem in ds:
    expected_vr = dictionary_VR(elem.tag)
    if expected_vr and elem.VR != expected_vr:
        print(f"Tag {elem.tag} ({elem.name}) has VR {elem.VR}, expected {expected_vr}")
```

This script will print out any tag whose VR does not match the standard. Similarly, one can check VM by comparing `len(elem.value)` to the defined multiplicity.

In practice, checks like that are run over all datasets to find structural errors (wrong VRs, too many values, etc.). Dedicated off-the-shelf tools offer more robust validation than custom scripts. For example, DVTk provides verification utilities for DICOM files. Another widely-used utility is dicom3tools, including `dciodvfy` and `dcentvfy`. `dciodvfy` examines each file's attributes against the Information Object Definition for its SOP Class, checks encoding rules, and validates tags using a relevant data dictionary. It will report missing required tags, incorrect VRs, and values outside allowed ranges. These tools also validate the integrity of UIDs and other key identifiers, ensuring they follow the correct syntax and structure as defined by the DICOM standard. For example, `dciodvfy` explicitly checks that each element adheres to pre-specified standards or formats. It checks for issues such as illegal UID roots, improper formatting, and leading zeros in numeric components, helping to maintain consistency and prevent identifier-related errors across systems. It flags both errors and warnings. For instance, a zero in *SliceThickness* or *Rows* might trigger a warning since they are unlikely in normal imaging. The companion `dcentvfy` utility checks consistency across multiple files (e.g., requiring that all instances in a series share the same *SeriesInstanceUID*).

Best practices followed by Bridge2AI's Functional Genomics, Salutogenesis and Precision Public Health teams combine automated checks of the tags listed in Table 3 - using `pydicom`, DVTk, `dicom3tools` - along with expert data review.

If any tag is nonstandard (private or repeating unexpectedly), it is documented. Data managers often maintain "private tag dictionaries" or use tools like The Cancer Imaging Archive's (TCIA) `tagSniffer` [33] to inventory all tags present and compare to expected profiles.

## *Stage 5: Image Quality and Pixel Data Validation*

In addition to the metadata found in DICOM headers, DICOM *pixel data* must be checked for readability and plausibility. This can be a compute-intensive step, depending on data modalities and scale.

Each DICOM's pixel array should decode properly. In Python, `pydicom` provides the

`ds.pixel_array` feature. If decoding raises an error, it indicates a problem (missing or corrupted *PixelData*, unsupported compression, etc.). The snippet below illustrates using this feature.

```python
  # Pydicom codesnippet checking pixels are correctly encoded
from pydicom import dcmread
try:
     ds = dcmread('Example_image.dcm')
     arr = ds.pixel_array
     print(f"Loaded image with shape {arr.shape}")
except Exception as e:
     print("Failed to read pixel data:", e)
```

If an image fails to load[2], the dataset owner should verify the *Transfer Syntax* and check whether the pixel data field is truncated.

Routine quality checks include confirming that images' dimensions (Rows × Columns) match expectations and verifying that pixel value ranges are plausible. For instance, a CT image should have Hounsfield Units in a reasonable range (e.g., -1000 to +3000); a mammogram should not have values beyond the detector's bit depth. If an image slice is uniformly zero or has an unexpected constant value, it may indicate an acquisition error or prior segmentation mask instead of actual data.

Even well-maintained archived imaging collections typically contain a small fraction of unreadable files (usually around 0.1%–0.5%), due to data acquisition or packaging issues. This underscores the need for automated scripts to test a sample or all images by reading them.

When dealing with compressed image data, one must also verify that decompression produces the correct bit depth and that no data loss occurred, for example, by comparing a JPEG-compressed DICOM to its original uncompressed version, if available. **Appendix III provides example JPEG compression values for endoscopy data**.

Once images are read, image quality metrics can be computed and documented. Simple metrics include image histogram checks (e.g., no extreme outliers), signal-to-noise ratio estimates, and sharpness measures.

While such checks are modality-dependent, common and easy steps towards data plausibility checks include computing summary statistics (mean, standard deviation of pixel values) per image or per series; and flagging values outside expected ranges (perhaps set from historical data or scanner specifications). For example, Radiology QC tools often include checks for artifacts (e.g., "salt and pepper" noise, excessive motion blur). The example below illustrates reading pixel data with `pydicom`, from an example MRI slice, printing basic image statistics.

```python
  # Pydicom codesnippet illustrating reading pixel data from an MRI 'brain' slice
```

---

[2] Some compressed formats like JPEG2000 require additional libraries installed as `pydicom` plugins

```python
from pydicom import dcmread
import numpy as np

ds = dcmread('Example_brainMRI_slice1.dcm')
img = ds.pixel_array   # NumPy array of pixel intensities
print("Min:", np.min(img), "Max:", np.max(img), "Mean:", np.mean(img))
```

This snippet confirms that the image's numeric content is accessible. If the DICOM file uses lossless compression (JPEG-LS, etc.), `pydicom` transparently handles it. If not, one can convert the image array to a standard format (e.g. PNG, NIfTI - Neuroimaging Informatics Technology Initiative, a popular alternative to DICOM images for working with biomedical data) for further analysis. Ultimately, to ensure that pixel data is intact and in the format specified by the header, both the bits and their meaning must be verified.

## Stage 6: Metadata Consistency and Correctness

After acquisition of basic tags and primary pixel data validation discussed above, verifying that image metadata is correct and consistent is a critical step, as **for large imaging datasets, metadata provides the only way data can be discovered, accessed and reused.**

Table 4 outlines key steps, methods and considerations in metadata reliability processes.

In addition to the critical general DICOM tags discussed above in Table 3, each modality has specific DICOM tags that are important. Bridge2AI grand challenge teams collect critical modality-specific DICOM tags in standard way across studies.

The Precision Public Health and AI/ML for Clinical Care grand challenge teams collect **CT** data including DICOM tags *Reconstruction Diameter*, *Convolution Kernel*, *Scan Length*, *kVp*, *Tube Current*. These teams also collect **MRI** data with tags *Repetition Time*, *Echo Time*, *Flip Angle*, *Magnet Strength*, *Sequence Name*.

Precision Public Health collects **ultrasound** data with tags *Probe Frequency*, *Ultrasound Doppler Type*, *Cine Rate*, *Frame Laterality* (for 3D probes).

The Salutogenesis team has assembled pioneering **ophthalmic** imaging datasets [5] along with tags *Image Type* (color, red-free, infra-red), *Illumination Source*, *Acquisition Device Data*, *Corneal Markers* (if present).

Finally, all teams collecting **video** DICOM data include the tag *Frame Time Vector*, along with *Presentation Group* attributes (if overlays are present).

**Metadata values** in tags must be both internally consistent and biologically plausible. Examples of bad data include a *PatientAge* of "150Y", or a *StudyDate* after a patient's death date. Detecting such issues often requires applying domain knowledge. For instance, *PatientAge* should typically be two digits (e.g., "085Y" for 85 years) and agree with the date of birth and study date. If a study of a knee has an *ImageLaterality* of "R", ensure the patient has an anatomy where laterality makes sense (e.g., laterality may not be applicable to an amputee).

**Table 4.** Key steps for metadata consistency and plausibility checks.

| Step | Description | Tools/Methods | Key Considerations |
|---|---|---|---|
| Validation | Check conformance to SOP Class requirements | DICOM validation tools (e.g., DCMTK's `dcmchk`, dcm4che; `dciodvfy`) | Catch missing required attributes or out-of-range values |
| Anonymization Check | Identify and remove/pseudonymize PHI in DICOM headers | DicomCleaner, `pydicom` scripts | Record anonymization date and method in file audit trails |
| Harmonization | Map free-text/local codes to standards | Use DICOM vocabulary (e.g., Modality tag), map terms to standard codes (SNOMED-CT or RadLex) | Ensure consistency across studies |
| Completeness | Verify all expected series/frames were extracted | Check DICOM tag Images in Acquisition or PACS count | Log missing files and attempt re-transfer if needed |
| Cross-Referencing | Link images to clinical context | HL7 orders or reports, capture identifiers (Accession Number, Study Instance UID) | Check consistency with image DICOM data |

Standard metadata tags support interoperability, while fragmented formats across devices impede downstream research. Harmonizing and integrating datasets across Bridge2AI studies for downstream AI and ML applications relies on the steps listed above to ensure metadata consistency and plausibility. In practice, these steps are built out as part of a broader DICOM data management pipeline.

## *Stage 7. Pixel-level and metadata de-identification*

Steps need to be taken to preserve patient privacy when sharing medical imaging data (Figure 3). For example, medical imaging of the head (e.g., MRI, CT) can inadvertently reveal a patient's face. Advances in 3D rendering and face-recognition algorithms make it possible to reconstruct facial features from volumetric scans. For example, Schwarz *et al.* [34] demonstrated that an MRI head scan can be reconstructed to match a photograph of the same person with ~83% accuracy. Facial characteristics like eyes, nose, and mouth are highly distinctive and even features such as ears carry identifying information. High-resolution scans that include facial or skull anatomy pose a re-identification risk that need to be remedied. Non-brain head scans (e.g., CT of sinuses, skull X-rays) likewise may reveal facial structure. Even images not explicitly showing the face (such as ophthalmic retinal scans or eye OCTA) can carry biometric identifiers (e.g., iris patterns) or contain patient data in metadata.

**Thus pixel-level de-identification ("defacing") of any imaging modality that can expose a person's face is critical for privacy when sharing data[3].**

---

[3] Both U.S. and EU privacy laws require removal of identifiable images. Under HIPAA, full-face photographs and any comparable images are explicitly defined as direct identifiers that must be removed for de-identification. The U.S. Safe Harbor method requires scrubbing not only metadata, but also any

Unfortunately, defacing is a complex, compute-intensive and costly task to perform at scale. Tools, described below, provide free open-source steps in defacing workflows. However, large-scale DICOM data aggregation projects need to budget and provision cloud or other compute resources to support these workflows' execution at scale.

Defacing methods fall into two broad categories: **skull-stripping** (removing all non-brain tissue) and **face-specific de-facing/masking** (see Figure 3).

Skull-stripping tools like FMRIB Software Library (FSL) `BET` [35] or Analysis of Functional NeuroImages (AFNI) `3dSkullStrip` [36] remove the scalp and skull entirely, which incidentally removes the face but also removes skull bone and sometimes neck tissues. This may present loss of valuable data: de-identification needs to be balanced with analysis utility downstream.

---

image that contains identifiable face features. Similarly, the EU GDPR classifies facial images as biometric personal data needing special handling. In practice, any DICOM data intended for public release or research sharing must have facial features obscured or eliminated to meet HIPAA/GDPR requirements. For example, the open-data platforms Human Connectome Project and OpenNeuro mandate defacing of structural MRI scans.

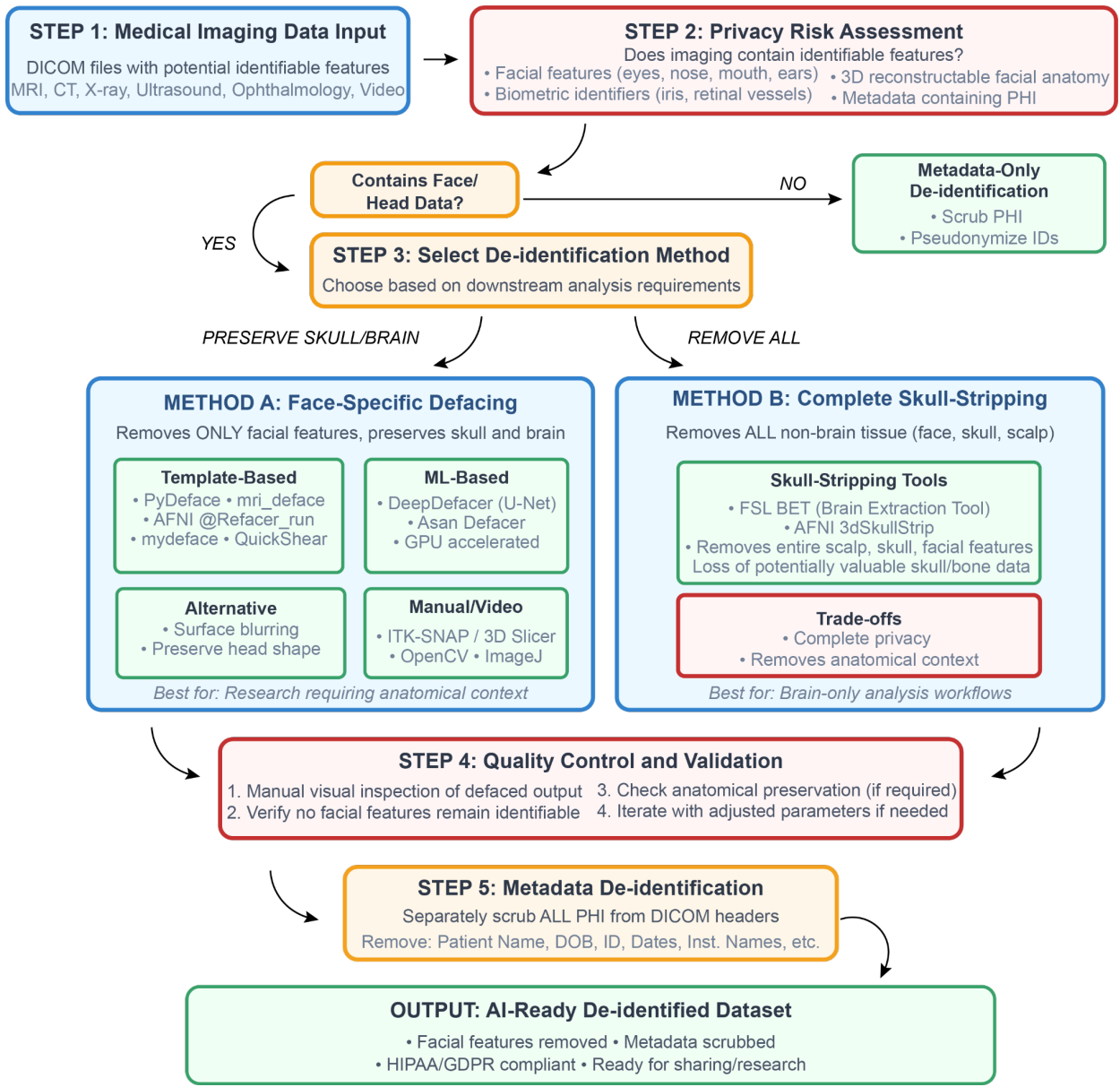

**Figure 3.** Overview of de-identification workflow. The initial medical imaging data is first assessed to see if it requires face/head de-identification or just metadata de-identification. If the images include face/head information that could compromise patient privacy, additional steps are required to remove facial features while potentially preserving skull and brain imaging.

In contrast, face-specific *de-facing* tools aim to remove only facial features (eyes, nose, mouth, ears, etc.) while preserving as much of the brain and head volume as possible. **Template-based** masking works by registering the image to a standard brain template (e.g., MNI152 [37]) and then applying a pre-defined binary face mask. For instance, `PyDeface` ([38]) aligns a subject's T1-weighted MRI to a template using FSL FLIRT ([39]) and then zeros out voxels in the facial mask region. Similarly, FreeSurfer's `mri_deface` ([40]) uses affine registration to fit a generic face mask to the data. BrainVoyager's DICOM Deface is a commercial tool that transforms a template-space face mask back into native space, similar to FreeSurfer. These methods automate de-facing with no manual intervention. They produce output where all voxels within the masked facial region are set to zero while the brain is unchanged.

Defacing workflows are often implemented in standard pipelines. For example, converting DICOM to NIfTI with `dcm2niix` or `dcm2bids` [41] is typically followed by running PyDeface on the structural images, as shown in the snippet below. Other example snippets can be found in the Neuroimaging Cookbook ([42]).

```
# a bash CLI tool for defacing using pydeface

$ pydeface input_T1w.nii.gz --outfile defaced_T1w.nii.gz
```

or equivalently in Python:

```python
from pydeface import PyDeface
defacer = PyDeface("Example_input_T1w.nii.gz")
defacer.run()  # produces input_T1w_defaced.nii.gz
```

The above snippet aligns the example input to a default template, applies the face mask, and saves a de-faced image. Other tools follow a similar script/API pattern, but often require FSL, Advanced Normalization Tools (ANTs), or other libraries.

Some approaches do *not* erase the face but blur or smear it. Milchenko & Marcus [43] use **surface blurring** by diffusing the face voxels to avoid sharp edges, preventing re-identification. Another approach, utilized by tools like QuickShear [44], computes a plane to shear off the face along a line through the head. These methods aim to hide identity while preserving more contextual head shape. However, template masking remains more common in neuroimaging pipelines.

Recent work uses **Machine Learning (ML) methods** like convolutional neural nets to segment and remove facial features. For example, Jeong *et al.* [45] employ a 3D U-Net to identify eyes, ears, nose and blurs only those surfaces; the code is available on GitHub. Likewise, DeepDefacer [46] is a U-Net trained on T1 MRI scans to generate a mask for facial regions. These ML methods can be faster on GPUs and may generalize across scans, but require a model training budget. They typically still zero out the identified face voxels, similar to template methods, discussed above.

If automatic de-facing fails or is unsuitable, **manual masking** de-facing tools could still facilitate a robust process. Image editing software like 3D Slicer or ITK-SNAP [47] can be used to paint a mask over the facial region. After segmentation, those voxels are zeroed or blurred. This trades off fine-grained control at the cost of extra work, but could also potentially offset cloud compute costs. Similarly, generic image tools (Photoshop, ImageJ) can blur or cover faces in 2D projections. For video sequences (e.g., DICOM ultrasound loops or endoscopy), face detection (e.g., OpenCV [48] Haar cascades) can identify faces frame-by-frame and mask or blur them as shown in the snippet below.

```python
# Python code snippet to blur a detected face in a video frame using OpenCV:
import cv2
face_cascade = cv2.CascadeClassifier('haarcascade_frontalface_default.xml')
cap = cv2.VideoCapture('Example_input_video.dcm')  # read DICOM video frames
```

```
while True:
    ret, frame = cap.read()

    if not ret: break
    gray = cv2.cvtColor(frame, cv2.COLOR_BGR2GRAY)
    faces = face_cascade.detectMultiScale(gray, 1.1, 4)

    for (x,y,w,h) in faces:
        face = frame[y:y+h, x:x+w]
        blurred = cv2.blur(face, (51,51))
        frame[y:y+h, x:x+w] = blurred

# (save or process further the de-faced frame) as needed
```

De-facing tools are not perfect. Template masks assume the subject's anatomy roughly matches the template space: unusual head shapes or implants can cause incomplete masking. Deep-learning defacers may generalize better but require GPUs for speed performance. Tools like QuickShear preserve more of the brain by removing only the front face, but can leave some features (eyes/ears) intact if not tuned. PyDeface's mask approach is conservative, often removing any potential facial voxels at the cost of cutting off tissue (e.g., portions of forehead or scalp). Manual checking of defaced output is always recommended. If facial features remain, one can iteratively adjust mask parameters or use alternate tools. Table 9 summarizes popular defacing tools, computational methods, and tools' licenses.

**Table 5:** Defacing tools, applicable modality and license.

| Tool/Pipeline | Supported Modality | Defacing Method | License/Availability |
|---|---|---|---|
| `PyDeface` | MRI (T1w) | Template-based masking (zeros out face) | Open-source (BSD) |
| FreeSurfer `mri_deface` | MRI (T1w) | Template mask (zeros out face) | Open-source (FreeSurfer) |
| AFNI `@Refacer_run` | MRI (T1w) | Template mask (often combined w/ skullstrip) | Open-source (AFNI) |
| QuickShear | MRI (T1w) | Cropping/shearing plane (remove front of head) | Open-source |
| DeepDefacer | MRI (T1w/T2w) | 3D U-Net segmentation & masking | Open-source (GitHub) |
| Asan Defacer | MRI, CT | 3D U-Net segments eyes/ears/nose, masks them | Open-source (GitHub) |
| `mydeface` [49] | MRI (T1w/FLAIR) | Similar to PyDeface; FSL FLIRT [39] mask | Open-source (BSD) |

| Skull stripping (FSL BET, AFNI 3dSkullStrip) | MRI, CT, PET | Removes all non-brain (full skull removal) | Open-source |
|---|---|---|---|
| ITK-SNAP / 3D Slicer | Any 3D volume (user choice) | Manual ROI paint or erode (face mask) | Open-source GUI |
| ImageJ / Fiji | 2D/3D images | Manual blur/crop of face region | Open-source |
| OpenCV (Python) [48] | Video, 2D images | Face detection + blur / pixelate | Open-source |
| OsiriX / Horos plugin | DICOM (multi-modality) | Built-in anonymization + face removal options | Commercial / Free |
| PixelMed DICOM Anonymizer | DICOM files | Metadata anonymization (no face masking) | Open-source |
| Manual (photos) | Photographs, videos | Crop/blur face using video/photo editors | — |

While some DICOM modalities like retinal images do not reveal the patient's face, they do capture internal eye structures. And ophthalmology scans of the eye can show eyelids and parts of the face. In practice, ophthalmic DICOM (like OCTA) usually needs only metadata scrub (names, dates), but high-resolution iris images *are* considered biometric under GDPR (the AAO notes iris scans require consent): care should be taken if sharing eye images of the face and iris.

**Other modalities:** Chest or abdominal CT/PET rarely include the face, so standard practice is to truncate images above the shoulders. Any image that includes a de-facing step. Similarly, whole-body scans that include face (PET/CT staging) should be cropped.

Note that defacing only removes pixel data. As part of a robust de-identification pipeline, all PHI fields (patient name, DOB, ID, etc.) must be stripped or pseudonymized separately using DICOM anonymizers noted in sections above.

# Conclusion

Effective, standards-based management of medical imaging data – including modalities such as brain MRIs, CT scans, X-rays, endoscopy and retinal scans, and ultrasound waveforms – is critical for advancing biomedical research. This article describes a practical, end-to-end framework for data management towards AI-ready medical imaging datasets by pairing DICOM data standard conformance with operationalized Standard Operating Procedures (SOPs) spanning data capture, verification, validation, and image de-identification. These SOPs are grounded in use cases and solutions arising across the NIH's Bridge2AI Clinical Care, Salutogenesis, Functional Genomics, and Precision Public Health grand challenges.

We supplement existing DICOM standard documentation with practical examples and code snippets, illustrating domain- and modality-specific DICOM file configurations. We cover the use of DICOM for representing multiple imaging modalities: static imaging data such as from MRIs, as well as videos such as endoscopies. We leverage standard tools to ensure data reliability – completeness, conformance, and biological plausibility – in the context of DICOM file integrity verification, tag validation, and pixel data quality checks, placing these tools in the context of a multi-stage data management strategy with modality-specific workflows. Anticipating potential use of AI and ML for adversarial re-identification of medical imaging datasets (e.g. head and upper-torso scans), we review existing methods for respective DICOM data de-identification.

The data management strategy outlined in this article helps to overcome the complexities of multi-institutional collaboration, site-specific data variations and ultimately provide downstream AI-ready medical imaging datasets.

# APPENDIX I

Accurate and reproducible management of medical imaging data begins at capture. By documenting imaging sources, using standardized DICOM toolkits and protocols, and documenting and verifying metadata quality early, data managers and researchers can avoid the inconsistencies that hinder downstream data use and interoperability. **Below is an example of a Data Landscape Assessment Questionnaire** capturing the four key dimensions of DICOM imaging data sources, discussed in the section entitled "Identifying and Documenting DICOM Imaging Sources".

Additionally, a Google form instantiation of the questionnaire can be found [here](here) and a respective editable form is [here](here).

---

**1. Project and Contact Information**
Project Name:
Lead Researcher:
Contact Information:
Date of Template Completion:

**2. Data Source Information**
Data Source Name:
Data Source Type: (e.g., Local PACS, Cloud Storage, Research Database)
Location/Institution:
System Contact Person:
Access Credentials Required: (Yes/No; if yes, specify what credentials or access procedures are required)

**3. Image Modality Details**
Required Modalities:
For each modality, please provide a brief explanation of the modality's relevance to the study

**4. Data Extraction Criteria**
Start Date:

End Date:
Reason for Date Range: (e.g., to capture baseline data, specific events, or treatment timelines)
Expected Total Data Size (if known):
Average Size per Study: (if available)

**5. Data Characteristics and Requirements**
Resolution Requirements: (e.g., High-resolution for analysis, Low-resolution for quicker loading)
File Format Requirements: (e.g., DICOM, JPEG for non-DICOM purposes)
Data Quality Checks:
Compliance with DICOM Standards: (Yes/No; additional details)
Reliability Requirements: (specify any specific requirements for checksum validation or metadata verification)

**6. Additional Notes on Image Metadata and Anonymization**
De-identification or Anonymization Needed: (Yes/No; specify level of anonymization required if necessary)
Specific Metadata Requirements: (e.g., fields needed, such as patient ID, date, modality, or institution)

**7. Data Transfer and Storage**
Storage Location for Extracted Data: (e.g., local drive, cloud storage)
Transfer Method: (e.g., secure FTP, encrypted storage device)
Estimated Duration for Extraction and Transfer:

**8. Approval and Compliance**
Ethics Approval or Institutional Review Board (IRB): (Yes/No; if Yes, specify details or documentation required)
Compliance with Data Privacy Regulations: (e.g., HIPAA, GDPR)

---

# APPENDIX II

Within Bridge2AI, the Salutogenesis grand challenge team carefully reviews DICOM files for conformance with published DICOM standards. This requires human review for each of the relevant tag values: e.g., for a tag Manufacturer, both the code and device name were checked to confirm a Manufacturer match). The Salutogenesis team identified a set of NEMA compliant versions of the DICOM files from each protocol (7 devices, 12 protocols). Table AII.1 lists the DICOM tags (SOP classes) used by the team for ophthalmology data extraction.

**Table AII.1.** Ophthalmology DICOM data extraction and conversion standards used by the Bridge2AI Salutogenesis grand challenge team [37].

| Converter Tasks | Examples | | |
|---|---|---|---|
| | Image Modality | SOP CLASS UID | SOP CLASS UID name |

| Evaluation of NEMA compliance (Information object definitions by DICOM Standards Committee within NEMA - refer to Figure 1) | OCT, OCTA | 1.2.840.10008.5.1.4.1.77.1.5.4 | Ophthalmic Tomography Image Storage |
|---|---|---|---|
| | CFP, FAF, IR, OCT, OCTA | 1.2.840.10008.5.1.4.1.77.1.5.1 | Ophthalmic Photography 8 Bit Image Storage |
| | OCTA | 1.2.840.10008.5.1.4.1.77.1.5.8 | B-scan Volume Analysis Storage |
| | OCTA | 1.2.840.10008.5.1.4.1.66.8 | Height Map Segmentation Storage |
| | OCTA | 1.2.840.10008.5.1.4.1.77.1.5.7 | En Face Image Storage |
| Incorporation of missing information | Pixel spacing information, manufacturer, etc. | | |
| Removal of demographic information | Patient Name, Birthdate, Sex/Gender, Race/Ethnicity, etc. | | |
| Harmonization of terms | Image Type, Anatomic Region Sequence, Manufacturer, Model Name, etc. | | |

The Salutogensis team wrote scripts to convert DICOM data exported from input devices to the compliant DICOM files, using `pydicom`. The data-capture scripts encode rules that are protocol and device specific (e.g., specific to Macular Cube 512 x 128 OCT protocol from a specific company). The team collaborated with many of the major manufacturers of color fundus photography equipment, OCT, and OCTAs to get manufacturer specific information as needed (e.g., filling in missing information on tags not provided by default in device outputs).

# APPENDIX III

The Bridge2AI Precision Public Health grand challenge team acquires DICOM-compliant video sequences from endoscopy, fluoroscopy, ultrasound, microscope, overhead operating camera modalities. Video DICOM data is stored in a PACS-compatible server, allowing for support of multi-frame DICOM objects. The team utilizes three storage setups, each with security, privacy benefits and challenges (e.g., cost, setup, governance compliance).

Each DICOM video data sequence follows Object Identification (IOD) specification defining the structure, data elements, and attributes of the image. The DICOM Service Object Pair (SOP) Class is the combination of an IOD and associated DICOM Service (e.g., storage, printing, query/retrieve). The SOP defines how a particular IOD is used. DICOM SOP Classes specify standard compression formats, metadata storage, interoperability parameters between medical imaging systems, and dynamic/real-time imaging (e.g., endoscopy, fluoroscopy, microscopy) metadata. All three PACS-compatible storage solutions mentioned above support these SOP Classes.

**Table AIII.1.** Video DICOM SOP classes used by the Bridge2AI Precision Public Health Team

| SOP Class Name | SOP Class Unique Identifiers (UID) |
|---|---|
| Video Endoscopic Image Storage | 1.2.840.10008.5.1.4.1.1.77.1.1.1 |
| Video Microscopic Image Storage | 1.2.840.10008.5.1.4.1.1.77.1.2.1 |
| Video Photographic Image Storage | 1.2.840.10008.5.1.4.1.1.77.1.4.1 |

A modality-specific and suitable choice of video and audio compression codecs is important for cost and quality trade-offs. Tables AIII.2 - AIII.6 provide a mapping between DICOM supported compression formats and data modalities used by the Precision Public Health team.

Video DICOM provides options for *lossless* compression to reduce file size without sacrificing image quality. The most common lossless compression methods used by the team are listed in Table AIII.2.

**Table AIII.2.** Video DICOM lossless compression formats along with applicable modalities generated by the Bridge2AI Precision Public Health Team

| Compression Type | DICOM Transfer Syntax (UID) | Compression Ratio | Applicable Modalities | Pros | Cons |
|---|---|---|---|---|---|
| MJPEG 2000 Lossless | 1.2.840.10008.1.2.4.92 | ~2:1 | Ultrasound, Medical video (e.g., fluoroscopy, endoscopy, surgical video), cine MRI | High quality, fully lossless (reversible wavelet compression) | High computational cost/storage needs, not as efficient as JPEG 2000 for high compression |
| JPEG 2000 Lossless | 1.2.840.10008.1.2.4.91 | ~2:1 | High-resolution medical images & video sequences | Ideal for static images and video sequences | High computational cost/storage needs |
| H.264 Near-Lossless (AVC) | 1.2.840.10008.1.2.4.103 | ~2:1 - 4:1 | Endoscopy, surgery, fluoroscopy | Efficient, widely supported | Requires modern PACS |

Video DICOM *lossy* compression reduces the file size, which is important for efficient storage and transmission, in certain scenarios. The types of lossy compression suitable for respective data modalities are listed in Table AIII.3.

**Table AIII.3:** Video DICOM lossy compression formats along with applicable modalities generated by the Bridge2AI Precision Public Health Team

| Compression Type | DICOM Transfer Syntax (UID) | Compression Ratio | Applicable Modalities | Pros | Cons |
|---|---|---|---|---|---|
| JPEG Lossy | 1.2.840.10008.1.2.4.50 | ~10:1 to 20:1 | Medical Images (static), Video Frames | Widely supported, good quality at moderate compression | Loss of quality, visible artifacts at high compression |
| MJPEG-2 Lossy | 1.2.840.10008.1.2.4.93 | ~50:1 to 100:1 | Video (e.g., endoscopy, surgery) | High compression ratio, reasonable quality | Lower video quality, some artifacts |
| MPEG-4 H.264 Lossy | 1.2.840.10008.1.2.4.103 | ~50:1 to 100:1 | Streaming video, surgical video | Excellent compression efficiency, high quality | Loss of quality, noticeable compression artifacts at low bitrates |
| H.265/ High-Efficiency Video Coding (HVEC) Lossy | 1.2.840.10008.1.2.4.104 | ~2.5:1 - 5:1 | 4K endoscopy, cine MRI, angiography | Best compression ratio, efficient | Limited support in older viewers |
| JPEG 2000 Lossy | 1.2.840.10008.1.2.4.50 | ~10:1 to 50:1 | Medical video with acceptable quality loss | Better compression than JPEG, less visible artifacts | Larger file size than MPEG/H.264, slower compression |

In certain applications, audio channels should accompany video ones. DICOM allows embedded audio. The Bridge2AI Precision Public Health Team uses the audio formats listed in Table AIII.4.

**Table AIII.4.** Audio DICOM formats to accompany video files generated by the Bridge2AI Precision Public Health Team

| Audio Format | Compression Type | DICOM Video Formats | Pros | Cons |
|---|---|---|---|---|
| WAV | Uncompressed | MJPEG, MPEG-4, H.264 | No loss of quality, ideal for medical procedures requiring high fidelity | Larger file sizes, computationally intensive |
| MP3 | Lossy compression | MPEG-4, H.264, HEVC | Smaller file sizes, widely compatible | Loss of quality at higher compression, audible artifacts |
| AAC | Lossy compression | H.264, HEVC, MPEG-4 | Efficient compression, better quality at lower bitrates | Loss of quality at higher compression, compatibility issues |
| LPCM (Linear PCM) | Uncompressed | MPEG-4, H.264, JPEG 2000 | High-quality audio, no loss of fidelity, ideal for critical procedures | Large file size, higher storage and bandwidth requirements |
| AC3 (Dolby Digital) | Lossy Compression | MPEG-4, H.264, HEVC | Support for multi-channel audio (5.1 surround), efficient compression | Potential loss of quality, compatibility limitations |
| MPEG-1 Layer II | Lossy Compression | MPEG-4, MJPEG, H.264 | Good quality at low bitrates, widely supported | Loss of quality, not as efficient as newer codecs like AAC |

Consistent frames rates, resolution, and compression standards must be maintained for reproducibility. Table AIII.5 and Table AIII.6 list frame rates options for lossless and lossy compression respectively.

**Table AIII.5.** Common lossless compression frame rates for video DICOM formats generated by the Bridge2AI Precision Public Health Team

| **Estimated MJPEG 2000 Lossless Bitrates** | | |
|---|---|---|
| *Resolution* | *Frame Rate* | *Bitrate (Mbps) - Approximate values for 4:4:4, 10-bit depth* |
| HD (1920×1080) | 24 fps | ~500-800 Mbps |
| HD (1920×1080) | 60 fps | ~1.2-1.5 Gbps |
| 2K (2048×1080) | 24 fps | ~600-900 Mbps |
| 2K (2048×1080) | 60 fps | ~1.3-1.8 Gbps |
| 4K (4096×2160) | 24 fps | ~2-3 Gbps |
| 4K (4096×2160) | 60 fps | ~4-6 Gbps |

**Table AIII.6.** Common lossy compression frame rates for video DICOM formats generated by the Bridge2AI Precision Public Health Team

| Estimated MJPEG-2 Lossy Bitrates | | |
|---|---|---|
| *Resolution* | *Frame Rate* | *Bitrate (Mbps) - Approximate values for 4:4:4, 10-bit depth* |
| HD (1920×1080) | 24 fps | ~100-150 Mbps |
| HD (1920×1080) | 60 fps | ~200-300 Mbps |
| 2K (2048×1080) | 24 fps | ~150-250 Mbps |
| 2K (2048×1080) | 60 fps | ~300-500 Mbps |
| 4K (4096×2160) | 24 fps | ~300-500 Mbps |
| 4K (4096×2160) | 60 fps | ~600-900 Mbps |

# APPENDIX IV

*Acronym Glossary*

# A

**AAC** – Advanced Audio Coding: Lossy audio compression format commonly used in video DICOM files

**AAO** – American Academy of Ophthalmology: Professional organization providing clinical guidelines

**AFNI** – Analysis of Functional NeuroImages: Open-source software suite for neuroimaging analysis, including skull-stripping tools

**AI** – Artificial Intelligence: Computer systems capable of performing tasks requiring human-like intelligence

**AI-READI** – Artificial Intelligence Ready and Equitable Atlas for Diabetes Insights: Bridge2AI dataset of standards-compliant DICOM retinal imaging

**ANTs** – Advanced Normalization Tools: Software toolkit for image registration and segmentation

**API** – Application Programming Interface: Set of protocols for building and integrating software applications

**AVC** – Advanced Video Coding

# B

**BET** – Brain Extraction Tool: FSL utility for removing non-brain tissue from neuroimages

**BIDS** – Brain Imaging Data Structure: Standard for organizing neuroimaging data

**Bridge2AI** – Bridge to Artificial Intelligence: NIH-funded program creating standardized, AI-ready biomedical datasets

**BSD** – Berkeley Software Distribution: Open-source software license type

# C

**C-FIND** – DICOM service command for querying imaging archives

**C-MOVE** – DICOM service command for retrieving images from archives

**C-STORE** – DICOM service command for storing images to archives

**CFP** – Color Fundus Photography: Retinal imaging modality capturing color images of the eye's interior

**CHoRUS** – Clinical Care Healthcare outcomes Research: Bridge2AI grand challenge focused on ICU data

**CLI** – Command Line Interface: Text-based interface for executing programs

**CNN** – Convolutional Neural Network: Deep learning architecture commonly used for image analysis

**CT** – Computed Tomography: Medical imaging modality using X-rays to create cross-sectional images

**CSV** – Comma-Separated Values: Plain text file format for tabular data

# D

**DCMTK** – DICOM Toolkit: Open-source C/C++ library implementing DICOM standards

**DICOM** – Digital Imaging and Communications in Medicine: International standard (ISO 12052) for medical imaging data

**DICOMweb** – Web-based DICOM services (QIDO-RS, WADO-RS, STOW-RS) for RESTful access to imaging data

# F

**FAF** – Fundus Autofluorescence: Retinal imaging technique highlighting metabolic activity

**FAIR** – Findable, Accessible, Interoperable, and Reusable: Principles for scientific data management

**FDA** – U.S. Food and Drug Administration: Federal agency regulating medical devices and drugs

**FHIR** – Fast Healthcare Interoperability Resources: HL7 standard for healthcare data exchange

**FLIRT** – FMRIB's Linear Image Registration Tool: FSL tool for aligning brain images

**fps** – frames per second: Measure of video temporal resolution

**FSL** – FMRIB Software Library: Comprehensive neuroimaging analysis toolkit

**FTP** – File Transfer Protocol: Standard for transferring files over networks

# G

**Gbps** – Gigabits per second: Data transfer rate measurement

**GDCM** – Grassroots DICOM: Open-source library for reading and writing DICOM files

**GDPR** – General Data Protection Regulation: EU privacy and data protection law

**GPU** – Graphics Processing Unit: Specialized processor for parallel computing tasks

**GUI** – Graphical User Interface: Visual interface for interacting with software

# H

**H.264/AVC** – Advanced Video Coding: Widely-used video compression standard (MPEG-4 Part 10)

**H.265/HEVC** – High Efficiency Video Coding: Advanced video compression offering better compression than H.264

**HD** – High Definition: Video resolution of 1920×1080 pixels

**HEVC** – High Efficiency Video Coding: See H.265

**HIPAA** – Health Insurance Portability and Accountability Act: U.S. law protecting patient health information

**HL7** – Health Level Seven: International standards organization for healthcare information exchange

# I

**ICU** – Intensive Care Unit: Hospital department for critically ill patients

**IHE** – Integrating the Healthcare Enterprise: Initiative promoting coordinated use of healthcare IT standards

**IOD** – Information Object Definition: DICOM specification defining data structure for each modality

**IR** – Infrared: Retinal imaging wavelength for visualizing deeper eye structures

**IRB** – Institutional Review Board: Committee reviewing research ethics

**ISO** – International Organization for Standardization: Global standards-setting body

**ITK** – Insight Toolkit: Open-source library for medical image processing and analysis

# J

**JPEG** – Joint Photographic Experts Group: Common image compression standard

**JSON** – JavaScript Object Notation: Lightweight data interchange format

# K

**kVp** – Peak Kilovoltage: X-ray tube voltage setting affecting image contrast and penetration

# L

**LPCM** – Linear Pulse Code Modulation: Uncompressed audio encoding method

# M

**Mbps** – Megabits per second: Data transfer rate measurement

**MD5** – Message Digest Algorithm 5: Cryptographic hash function for data integrity verification

**MJPEG** – Motion JPEG: Video compression treating each frame as separate JPEG image

**ML** – Machine Learning: Subset of AI focused on learning from data

**MNI** – Montreal Neurological Institute: Developer of standardized brain templates (e.g., MNI152)

**MP3** – MPEG-1 Audio Layer 3: Common lossy audio compression format

**MPEG** – Moving Picture Experts Group: Standards organization for audio/video compression

**MRI** – Magnetic Resonance Imaging: Medical imaging using magnetic fields and radio waves

# N

**NEMA** – National Electrical Manufacturers Association: Organization maintaining DICOM standards

**NIfTI** – Neuroimaging Informatics Technology Initiative: Neuroimaging file format alternative to DICOM

**NIH** – National Institutes of Health: U.S. federal biomedical research agency

**NIST** – National Institute of Standards and Technology: U.S. agency setting technology standards

# O

**OCT** – Optical Coherence Tomography: High-resolution retinal imaging technique

**OCTA** – Optical Coherence Tomography Angiography: OCT-based retinal blood vessel imaging

**OHDSI** – Observational Health Data Sciences and Informatics: Collaborative for healthcare data standardization

**OMOP** – Observational Medical Outcomes Partnership: Common data model for healthcare research

# P

**PACS** – Picture Archiving and Communication System: Medical imaging storage and retrieval system

**PET** – Positron Emission Tomography: Nuclear medicine imaging technique

**PHI** – Protected Health Information: Individually identifiable health data protected under HIPAA

**PNG** – Portable Network Graphics: Lossless image compression format

**PS3.x** – Part x of DICOM Standard: Notation for DICOM standard sections (e.g., PS3.3 for Information Object Definitions)

# Q

**QA** – Quality Assessment: Systematic evaluation of data or process quality

**QIDO** – Query based on ID for DICOM Objects: DICOMweb service for searching imaging studies

# R

**REST/RESTful** – Representational State Transfer: Web service architectural style

**RGB** – Red Green Blue: Additive color model for digital images

**ROI** – Region of Interest: Selected portion of image for analysis

**RSNA** – Radiological Society of North America: Professional radiology organization

# S

**SHA** – Secure Hash Algorithm: Cryptographic hash functions (e.g., SHA-256) for data integrity

**SNOMED** – Systematized Nomenclature of Medicine: Comprehensive medical terminology system

**SOP** – Service-Object Pair (in DICOM context): Combination of Information Object Definition and DICOM service

**SOP** – Standard Operating Procedure (in data management context): Documented process for performing operations consistently

**STOW** – Store Over the Web: DICOMweb service for uploading images

# T

**TCIA** – The Cancer Imaging Archive: Public repository of cancer imaging data

**TE** – Echo Time: MRI parameter measuring time between excitation and signal detection

**TR** – Repetition Time: MRI parameter measuring time between successive pulse sequences

# U

**U-Net** – U-shaped convolutional neural network architecture for image segmentation

**UID** – Unique Identifier: Globally unique alphanumeric string identifying DICOM entities

**URL** – Uniform Resource Locator: Web address

# V

**VM** – Value Multiplicity: DICOM specification for number of values in a data element

**VNA** – Vendor Neutral Archive: Multi-vendor medical imaging storage system

**VR** – Value Representation: DICOM data type specification (e.g., DS=Decimal String, UI=Unique Identifier)

# W

**WADO** – Web Access to DICOM Objects: DICOMweb service for retrieving images

**WAV** – Waveform Audio File Format: Uncompressed audio format

# X

**XLSX** – Microsoft Excel Open XML Spreadsheet: Spreadsheet file format

**XNAT** – Extensible Neuroimaging Archive Toolkit: Open-source platform for managing and sharing medical imaging data

## Numbers & Symbols

**2D** – Two-Dimensional: Flat imaging (e.g., X-ray)

**2K** – Video resolution of approximately 2048 horizontal pixels

**3D** – Three-Dimensional: Volumetric imaging (e.g., CT, MRI)

**4K** – Video resolution of approximately 4096 horizontal pixels (Ultra HD)